\documentclass[aps,prb,onecolumn,showpacs,showkeys,preprintnumbers,superscriptaddress]{revtex4}
\usepackage[T1]{fontenc}
\usepackage{subeqn}
\usepackage[latin9]{inputenc}
\usepackage{graphicx}
\usepackage{amssymb}
\usepackage{epsfig}
\usepackage{epic}
\usepackage{pifont}
\usepackage{fancyhdr}
\usepackage{verbatim}
\usepackage{slashed}
\usepackage{subfigure}
\usepackage{url}
\usepackage{multirow}
\usepackage{color}

\providecommand{\boldsymbol}[1]{\mbox{\boldmath $#1$}}

\newcommand{\ee}{\end{equation}}
\newcommand{\be}{\begin{equation}}
\newcommand{\bea}{\begin{eqnarray}}
\newcommand{\eea}{\end{eqnarray}}

\newcommand{\eu}{{\rm e}}
\newcommand{\ii}{{\rm i}}
\newcommand{\de}{{\displaystyle\rm\mathstrut d}}
\newcommand{\sgn}{{\rm sgn}}
\newcommand{\sn}{{\rm sn}}
\newcommand{\cn}{{\rm cn}}
\newcommand{\dn}{{\rm dn}}
\newcommand{\tJ}{{\tilde{J}}}

\usepackage[english]{babel}
\makeatother

\begin{document}


\title{Modular invariance in the gapped $XYZ$ spin $1 \over 2$ chain}

\preprint{MIT-CTP 4414}

\author{Elisa Ercolessi}
\affiliation{Department of Physics, University of Bologna and I.N.F.N., Sezione di Bologna, Via Irnerio 46, 40126, Bologna, Italy}

\author{Stefano Evangelisti}
\affiliation{Department of Physics, University of Bologna and I.N.F.N., Sezione di Bologna, Via Irnerio 46, 40126, Bologna, Italy}

\author{Fabio Franchini}
\affiliation{Department of Physics, Massachusetts Institute of Technology, Cambridge, Massachusetts 02139, U.S.A.}
\affiliation{SISSA and I.N.F.N, Via Bonomea 265, 34136, Trieste, Italy}
\email{fabiof@mit.edu}

\author{Francesco Ravanini}
\affiliation{Department of Physics, University of Bologna and I.N.F.N., Sezione di Bologna, Via Irnerio 46, 40126, Bologna, Italy}

\begin{abstract}
We show that the elliptic parametrization of the coupling constants of the quantum $XYZ$ spin chain can be analytically extended outside of their natural domain, to cover the whole phase diagram of the model, which is composed of 12 adjacent regions, related to one another by a spin rotation. This extension is based on the modular properties of the elliptic functions and we show how rotations in parameter space correspond to the double covering $PGL(2,\mathbb Z ) $  of the modular group, implying that the partition function of the $XYZ$ chain is invariant under this group in parameter space, in the same way as a Conformal Field Theory partition function is invariant under the modular group acting in real space. 
The encoding of the symmetries of the model into the modular properties of the partition function could shed light on the general structure of integrable models.
\end{abstract}



\maketitle

\section{Introduction}

The identification of modular properties in integrable theories has always represented  a progress in our understanding of the underlying physical and mathematical structures of the model. Modular transformations can appear in a physical model in many different ways. For example, in two-dimensional Conformal Field Theory (CFT) the modular invariance of partition functions on a toroidal geometry \cite{Cardy86} has led to the classification of entire classes of theories \cite{CIZ} and, in many cases, has opened the way to their higher genus investigation, which is a very important tool in applications ranging from strings to condensed-matter physics.
In a different context, modular transformations are at the root of electromagnetic duality in (supersymmetric) gauge theories. \cite{SeibergWitten} In this case the modular transformations act on the space of parameters of the theory, rather than as space-time symmetries.

In integrable two-dimensional lattice models modular properties were crucial in a Corner Transfer Matrix (CTM) approach \cite{Baxter} to get the so-called highest weight probabilities \cite{JimboMiwa,BauerSaleur}. 
More recently, the problem of computing bipartite entanglement Renyi entropies (or their von Neumann limit) in integrable spin chains has been related to the diagonalization of the CTM of the corresponding two-dimensional (2D) lattice integrable model. The expressions for these quantities often present a form that can be written in terms of modular functions, like the Jacobi $\theta$ functions, suggesting links with a modular theory in the space of parameters.
In the $XY$ chain, for example, modular expressions have been found for the entanglement entropy in Ref. \onlinecite{Franchini}.
Similar modular expressions have been obtained in the calculation \cite{Evangelisti, Ercolessi2011} of entanglement entropies in the $XYZ$ spin chain, notoriously linked in its integrability to the eight-vertex lattice model. \cite{Baxter} The $XYZ$ model is one of the most widely studied examples of a fully interacting but still integrable theory,  playing the role of a sort of ``father'' of many other very interesting and useful spin chain models, like the Heisenberg model and its $XXZ$ generalization. Here we would like to start the investigation of its modular symmetries, which --let us stress it again-- are not space-time symmetries, like in CFT, but symmetries in the space of parameters.
We believe that the exploiting of modular properties hidden behind the Baxter reparametrization of the $XYZ$ chain can lead, similarly to the aforementioned examples of conformal and gauge theories, to new tools in the investigation of crucial properties of the $XYZ$ chain and its family of related models. 

In this paper we show that the $XYZ$ partition function is invariant under a set of transformations that implement a modular group covering the full plane of parameters of the model in a very specific way. This, in our opinion, prepares the way to the application of the modular group to the study of more complicated quantities, that we plan to investigate in the near future. For instance, this modular invariance may play an important role in a deeper understanding of the Renyi and von Neumann bipartite entanglement entropies, which are well known to be excellent indicators to detect quantum phase transitions. 
The paper is organized as follows. In Sec. \ref{sec:XYZsym} we recall basic facts about the Hamiltonian formulation of the $XYZ$ spin chain, the properties of its phase-space diagram and the symmetry transformations leading one region of the diagram into another.
In Sec. \ref{sec:Baxter} we recall Baxter's mapping of the eight-vertex model's parameters through elliptic functions and the relation between them and those of the $XYZ$ chain in the different parts of its phase diagram. Section \ref{sect:newparameters} is devoted to the analytical extension of Baxter solution by switching to new parameters that make the modular properties more evident. This shows how one of the phase-space regions can be mapped into the others by applying a modular transformation. The generators of these modular transformations are identified in the next section, Sec. \ref{sec:modular}, where the characterization of the actions of the full modular group is described. Our conclusions and outlooks are drawn in Sec. \ref{sec:conclusions}, while Appendix \ref{app:elliptic} collects some definitions and useful identities on elliptic functions.

\section{$XYZ$ chain and its symmetries}
\label{sec:XYZsym}

The quantum spin-$\frac{1}{2}$ ferromagnetic $XYZ$ chain can be described by the following Hamiltonian:
\begin{equation}
   \hat{H}_{XYZ} =
   -{\displaystyle \sum_{n}} \Big[ J_{x}\sigma_{n}^{x} \sigma_{n+1}^{x}
   + J_{y}\sigma_{n}^{y} \sigma_{n+1}^{y} + J_{z}\sigma_{n}^{z} \sigma_{n+1}^{z} \Big] \; ,
   \label{eq:XYZ1}
\end{equation}
where the $\sigma_{n}^{\alpha}$ ($\alpha=x,y,z$) are the Pauli matrices acting on the site $n$, the sum ranges over all sites $n$ of the chain and the constants $J_{x}$, $J_{y}$ and $J_{z}$ take into account the degree of anisotropy of the model. We assume periodic boundary conditions.

\begin{figure}
   \hskip-1cm\includegraphics[scale=0.7]{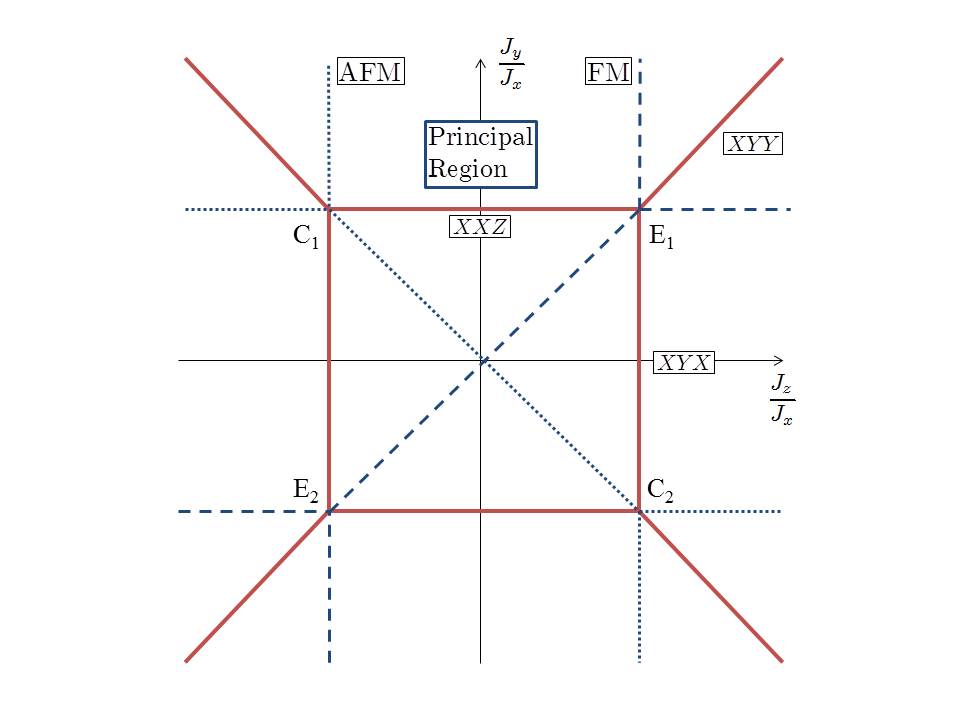}
   \caption{(Color online) Phase diagram of the $XYZ$ model in the $\left( {J_z \over J_x} , {J_y \over J_x} \right)$ plane. The critical $XXZ$, $XYX$, and $XYY$ chains are represented by bold (red) lines. Dashed (blue) lines are the continuations of the critical phases into the Ferro-/AntiFerro-Magnetic gapped chain (dashed/dotted), ending at infinity on a FM/AFM Ising model. There are four tricritical points where the three gapless phases meet: two of them, $C_{1,2}$, correspond to the isotropic AFM Heisenberg chain and  relativistic low-energy excitations (spinons), while the other two, $E_{1,2}$, give the isotropic FM chain, whose low-energy excitations have a quadratic dispersion relation (magnons).}
   \label{fig:phasediagram}
\end{figure}

The space of parameters $(J_x , J_y, J_z)$ forms a projective space, since an overall rescaling only changes the energy units. To visualize the phase diagram of the model we can single out one of them, say $J_x$, which, without loss of generality we take as positive, and switch to homogeneous coordinates:
\be
  \tJ_y \equiv {J_y \over J_x}  \qquad {\rm and} \qquad
  \tJ_z \equiv {J_z \over J_x} \; .
  \label{homcoord}
\ee
However, as $J_x$ varies, this mapping introduces artificial discontinuities when $J_x$ vanishes. The appearance of these singularities is unavoidable: it is just due to the choice of using a single two-dimensional map to describe the three-dimensional space of parameters and it is a known phenomenon in the study of projective spaces. We will soon deal with this branch discontinuities in discussing the symmetries of the model. Later on in Sec. \ref{sec:modular}, however, we will revert to the full three-dimensional parameter space in order to avoid these singularities and to show more clearly the action of the modular group.

In Fig. \ref{fig:phasediagram} we draw a cartoon of the phase diagram of the $XYZ$ model in the $\left( {J_z \over J_x}, {J_y \over J_x} \right) =(\tJ_z, \tJ_y) $ plane. We divide it into 12 regions, whose role will become clear as we proceed.
Let us first look at the $J_y=J_x$ line. Here we have the familiar $XXZ$ model and we recognize the critical (paramagnetic) phase for 
$\left| \tJ_z \right|<1$ (bold, red online, continuous line), the antiferromagnetic Ising phases for $\tJ_z <-1$ (dotted line) and the Ising ferromagnetic phase for $\tJ_z >1$ (dashed line).
The same physics can be observed for $J_y= -J_x$. There, one can rotate every other spin by 180° around the $x$-axis to recover a traditional $XXZ$ model. Note, however, that $J_z$ changes sign under this transformation and therefore the ferromagnetic and anti-ferromagnetic Ising phases are reversed.
The lines $J_z = \pm J_x$ correspond to a $XYX$ model, i.e. a rotated $XXZ$ model. Thus the phases are the same as before.
Finally, along the diagonals $J_y = \pm J_z$ we have a $XYY$ model of the form
\be
   \hat{H}_{XYY} =
   -{\displaystyle \sum_{n}} \left[ J_x \sigma_{n}^{x} \sigma_{n+1}^{x}
   + J_{y} \left( \sigma_{n}^{y} \sigma_{n+1}^{y} \pm \sigma_{n}^{z} \sigma_{n+1}^{z} \right) \right] \; .
   \label{HXYY}
\ee
Thus the paramagnetic phase is for $\left| \tJ_y \right| > 1 $ and the Ising phases for $\left| \tJ _y\right| <1$, with the plus sign ($J_y = J_z$) for Ising ferromagnet and the minus ($J_y = -J_z$) for the antiferromagnet.

In this phase diagram we find four {\it tricritical points} at $\left( \tJ_z  , \tJ_y  \right) = (\pm 1, \pm 1)$.
In the vicinity of each paramagnetic phase, in a suitable scaling limit, \cite{Evangelisti,Luther,John} the model renormalizes to a sine-Gordon theory, with $\beta^{2} = 8 \pi$ at the anti-ferromagnetic isotropic point and to a $\beta^{2} \rightarrow 0$ theory toward the ferromagnetic Heisenberg point. This is the reason why the critical nature of these tricritical points is quite different. Assuming $J_x >0$, at $C_{1}= \left( \tJ_z , \tJ_y \right) = (-1, 1)$ and $C_{2} = \left( \tJ_z , \tJ_y \right) = (1, -1)$ we have two conformal points dividing three equivalent (rotated) sine-Gordon theories with $\beta^{2} = 8 \pi$. At $E_{1} = \left( \tJ_z , \tJ_y \right) = (1, 1)$ and $E_{2} = \left( \tJ_z , \tJ_y \right) = (- 1, - 1)$, we have two $\beta^{2} = 0$ points which are no longer conformal, since the low-energy excitations have a quadratic spectrum there. The former points correspond to an antiferromagnetic Heisenberg chain at the Berezinkii-Kosterlitz-Thouless (BKT) transition, while the latter correspond to a Heisenberg ferromagnet at its first-order phase transition. The different nature of the tricritical points has been highlighted from an entanglement entropy point of view in Ref. \onlinecite{Ercolessi2011} .

For $J_x=J_y=\pm J_z$, we deal with an (anti)ferromagnetic Heisenberg model. As soon as some anisotropy is introduced, its original full SU(2) symmetry is broken: along any of the $XXZ$ lines a U(1) symmetry is left, whereas in the most general situation the residual $\mathbb{Z}_2$ symmetry group is generated by  $\pi$ rotations, $\prod_n \eu^{\ii (\pi / 2) \sigma_n^\alpha}$, $\alpha=x,y,z$, of every spin around one of the axes. In fact, as already observed, the partition function generated by Hamiltonian (\ref{eq:XYZ1}) is invariant under $\pi/2$ rotations ${\bf R}_\alpha = \prod_n \eu^{\ii (\pi / 4) \sigma_n^\alpha}$, since the effect of these transformations can be compensated by a simultaneous exchange of two couplings: $E_\alpha : J_\beta \leftrightarrow J_\gamma$ being $(\alpha, \beta, \gamma)$ a cyclic permutation of $(x,y,z)$. Thus, this $\mathbb{Z}_2$ symmetry of the Hamiltonian is enhanced to become the {\it symmetric group} (of permutations of three elements) ${\cal S}_3$ at the level of the partition function. 

The partition function shows an additional $\mathbb{Z}_2$ symmetry, generated by a $\pi$ rotation of every other spin around one of the axis ${\bf \Pi}_\alpha = \prod_n \sigma_{2 n}^\alpha$, which can be compensated by the reversal $P_\alpha$ of the signs of the couplings in the corresponding plane. 
The total symmetry group is generated by these six elements, which are however not all independent. There are only three independent generators that we choose: $P_x, E_x, E_y$. In fact, in choosing homogeneous coordinates (\ref{homcoord}), we cannot implement $P_y$ or $P_z$, since they would invert the sign of one between $\tJ_y$ and $\tJ_z$ , thus changing the parity of the signs of the couplings and hence inverting ferromagnetic and antiferromagnetic phases.

\begin{figure}
   \hskip-1cm\includegraphics[scale=0.7]{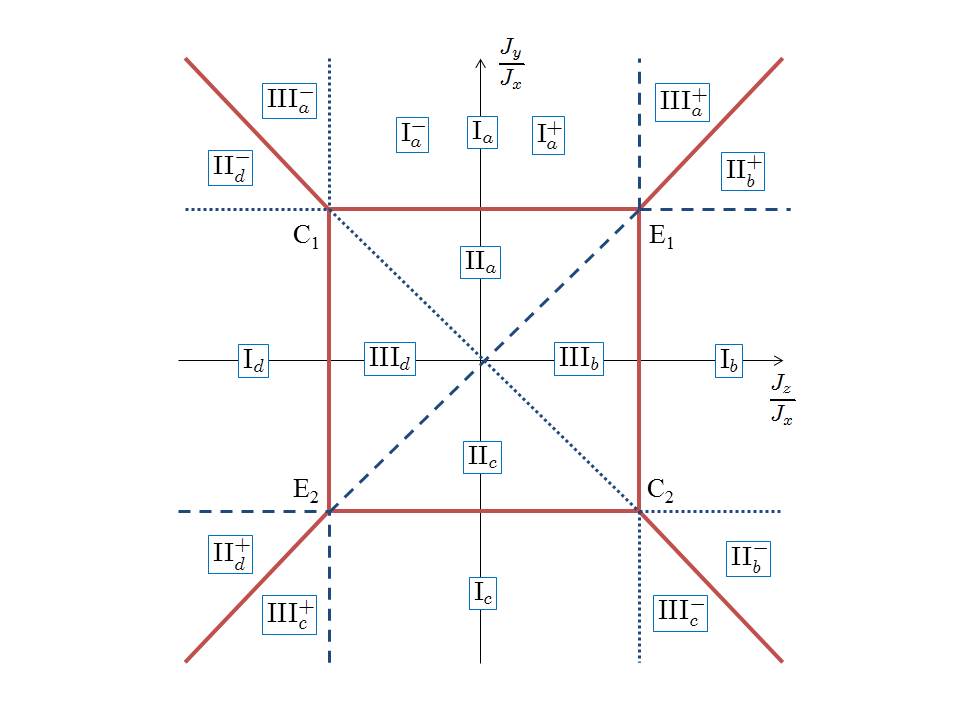}
   \caption{(Color online) The phase diagram of the $XYZ$ chain in Fig.~\ref{fig:phasediagram} is divided into 12 regions, named ${\rm I}_{a,b,c,d}$, ${\rm II}_{a,b,c,d}$, ${\rm III}_{a,b,c,d}$. Each region is further separated into the $^+$ half lying around one of the FM isotropic points $E_{1,2}$ and the $^-$ half closer to the AFM conformal points $C_{1,2}$.}
   \label{fig:phasediagramXYZ}
\end{figure}

We may look at the effect of these symmetries in the space of parameters $\{ J_x , J_y , J_z \}$. Explicitly, the generators act as
\begin{equation}
\begin{array}{l} 
P_x : \{ J_x , J_y , J_z \} \mapsto  \{ J_x , -J_y , - J_z \} \; , \\
E_x : \{ J_x , J_y , J_z \} \mapsto \{ J_x , J_z , J_y \} \; , \\
E_y : \{ J_x , J_y , J_z \} \mapsto \{ J_z , J_y , J_x \} \; . 
\end{array} 
\end{equation}
and relate the different regions of the phase diagram shown in Fig. \ref{fig:phasediagramXYZ}, named ${\rm I}_{a,b,c,d}$, ${\rm II}_{a,b,c,d}$, ${\rm III}_{a,b,c,d}$. In particular, $P_x$ corresponds to a $\pi$ rotation in the plane mapping, for example, region ${\rm I}_a $ onto  ${\rm I}_c$. Under this transformation, for instance, the half plane to the right of the antidiagonal $\tJ_z  = - \tJ_y$ is mapped into the half plane to the left of it.
The transformation $E_x$ instead corresponds to a reflection about the main diagonal, mapping e.g. region ${\rm I}_a $ onto  ${\rm I}_b$. Combined, these two transformations transform any of the four sectors of the plane lying in between the diagonal and the antidiagonal (that we distinguish with the letters $a,b,c,d$) one into the others. Inside each of these sectors, we will denote with superscripts $^+$ and $^-$ the subregions of $I,II,III$ touching the critical points of type $E$ and $C$ respectively, again as indicated in Fig. \ref{fig:phasediagram}.  For example ${\rm I}_a^+ =\{0<\tJ_z <1, \tJ_y>1\}$ and $ {\rm I}_c^-= \{0<\tJ_z <1, \tJ_y<-1\}$.
To see how $E_y$ works, let us start from  ${\rm I}_a^+ $. Now $E_y$ exchanges the role of $x$ and $z$, mapping a point $(\tJ_z , \tJ_y)\in {\rm I}_a^+ $ to the point with coordinates $\left( {J_x \over J_z} , {J_y \over J_z} \right) = \left( {1 \over \tJ_z} , {\tJ_y \over \tJ_z} \right)$,  which belongs to the zone in the first quadrant which we denote with ${\rm III}_a^+$. Naively, $E_y$ maps ${\rm I}_a^-$ into ${\rm III}_c^+$, but this would interchange the antiferromagnetic nature of the region into a ferromagnetic one, since this transformation would not preserve the parity of signs in the couplings. Thus, in order to project the three-dimensional space of parameter on the two-dimensional plane, we define $E_y \simeq E_y \circ P_y$ when acting on $J_z <0$, so that it maps ${\rm I}_a^-$ correctly into ${\rm III}_a^-$.  Similarly, one can see that the area ${\rm II}_a$  is reached starting from ${\rm I}_a$ via the $E_z$ exchange, which in turn can be written as $E_y\circ E_x\circ E_y$. In this way, we covered the whole area above the two diagonals. As said before, we can reflect about the diagonal acting with $E_x$, thus covering the whole region to the left of the antidiagonal quadrant. Finally,  the remaining half plane can be reached via a $\pi$ rotation $P_x$, as explained above.

In the following sections we will show that the transformations explained above are equivalent to the action of an extension of the modular group. By taking the principal region ${\rm I}_a$ as a reference starting point, in Sec. \ref{sec:modular} we collect the action of the different operators in Table \ref{transformationtable}.

\section{Baxter's solution}
\label{sec:Baxter}

The Hamiltonian (\ref{eq:XYZ1}) commutes with the transfer matrices of the zero-field eight vertex model \cite{Baxter} and this means that they can be diagonalized together and they share the same eigenvectors. 
Indeed, as shown by Sutherland, \cite{Sutherland} when the coupling constants of the $XYZ$ model are related to the parameters $\Gamma$ and $\Delta$ of the eight vertex model\footnote{We adopt the conventions of [\onlinecite{Baxter}] on the relation among $\Gamma,\Delta$ and the Boltzmann weights of the  8-vertex model.} at zero external field by the relations
\begin{equation}
   J_{x}:J_{y}:J_{z}=1:\Gamma:\Delta \; ,
   \label{eq:XYZ4bis}
\end{equation}
the row-to-row transfer matrix ${\bf T}$ of the latter model commutes with the Hamiltonian $\hat{H}$ of the former.
In the principal regime of the eight vertex model it is customary \cite{Baxter} to parametrize the constants $\Gamma$ and $\Delta$ in terms of elliptic functions,
\begin{equation}
   \Gamma = \frac{1+k\;\mbox{sn}^{2}(\ii \lambda; k)}{1-k\;\mbox{sn}^{2}(\ii \lambda;k)} \; ,
   \qquad\qquad
   \Delta = -\frac{\mbox{cn}(\ii \lambda;k)\;\mbox{dn}(\ii \lambda;k)}{1-k\;\mbox{sn}^{2}(\ii \lambda;k)} \; ,
   \label{eq:XYZ3bis}
\end{equation}
where $\mbox{sn}(z;k)$, $\mbox{cn}(z;k)$ and $\mbox{dn}(z;k)$ are Jacoby elliptic functions of parameter $k$, while $\lambda$ and $k$ are the argument and the parameter (respectively) whose natural regimes are 
\begin{equation}
   0 \le k \le 1\; , \qquad \qquad 0 \le \lambda \le K(k') \; ,
   \label{eq:XYZ4}
\end{equation}
$K(k')$ being the complete elliptic integral of the first kind (\ref{Kdef}) of argument $k'=\sqrt{1-k^{2}}$.

In light of (\ref{eq:XYZ4bis}), Baxter's parametrization encourages us to use the homogeneous coordinates (\ref{homcoord}), although this will introduce artificial discontinuities when we will extend it to the whole phase diagram. This is the aforementioned issue in using a two-dimensional map to describe a three-dimensional projective space. In Sec.\ref{sec:modular}, we will give an alternative prescription, which highlights the action of the modular group and acts directly on the three-dimensional parameter space, thus avoiding any discontinuity.
 
Equations (\ref{eq:XYZ4bis}) and (\ref{eq:XYZ3bis}) have the great limitation that, in the natural regime of the parameters $(\lambda,k)$, $\Delta \le -1$, $|\Gamma| \le 1$, which correspond to the antiferroelectric phase of the eight vertex model. Thus, Eq. (\ref{eq:XYZ4bis}) only covers region ${\rm I}_d$ of Fig. \ref{fig:phasediagramXYZ}. However, using the symmetries of the model and the freedom under the rearrangement of parameters, one can take the antiferroelectric $(\Gamma, \Delta)$ in (\ref{eq:XYZ3bis}) and use them in the whole phase diagram of the $XYZ$ chain, generalizing (\ref{eq:XYZ4bis}). The result of this procedure gives, when $\left| \tJ_y \right|<1$ and $\left| \tJ_z \right|<1$:
\begin{equation}
   \label{st11}
   \left\{
   \begin{array}{r c l}
      \Gamma & = &  {\displaystyle\frac 
      {\left| \tJ_z - \tJ_y \right|  - \left| \tJ_z + \tJ_y \right| }  
      {\left| \tJ_z - \tJ_y \right| + \left| \tJ_z + \tJ_y \right|}} \; , \\
      \Delta & = &    {\displaystyle-\frac{2}
     {\left| \tJ_z - \tJ_y \right| + \left| \tJ_z + \tJ_y \right|}} \, ,
   \end{array} \right.
   \qquad \qquad \left| \tJ_y \right|<1 \: \: {\rm and} \: \: \left| \tJ_z \right|<1 ,
\end{equation}
while for $\left| \tJ_y \right|>1$ or $\left| \tJ_z \right|>1$, we get:
\begin{equation}
   \label{st12}
   \left\{
   \begin{array}{r c l}
      \Gamma & = &  {\displaystyle\frac
      {\min\left[1,\left|\frac{\left|\tJ_z - \tJ_y\right| - \left|\tJ_z + \tJ_y\right|}{2}\right|\right]}
      {\max\left[1,\left|\frac{\left|\tJ_z - \tJ_y\right| - \left|\tJ_z + \tJ_y\right|}{2}\right|\right]}
      \cdot \mbox{sgn} \Big( \left|\tJ_z -\tJ_y\right| - \left|\tJ_z + \tJ_y\right| \Big)} \; ,\\
      \Delta & = &
      {\displaystyle-\frac{1}{2}\frac{\left|\tJ_z - \tJ_y\right| + \left|\tJ_z + \tJ_y\right|}
      {\max\left[1,\left|\frac{\left|\tJ_z - \tJ_y \right| - \left|\tJ_z + \tJ_ y \right|}{2}\right|\right]}} \; ,
   \end{array} \right.
   \qquad \qquad \left| \tJ_y \right|>1 \: \: {\rm or} \: \: \left| \tJ_z \right|>1 ,
\end{equation}
where, we remind, $\Gamma$ and $\Delta$ are defined by (\ref{eq:XYZ3bis}).
For instance in region ${\rm I}_d$, for $|\tJ_y| \le 1$ and $\tJ_z \le-1$, we recover $\tJ_y = \Gamma$ and $\tJ_z = \Delta$, while in region ${\rm I}_a$,  for $\tJ_y \ge 1$ and $|\tJ_z| \le 1$, we have $\tJ_y = - \Delta$ and $\tJ_z = -\Gamma$. In Fig. \ref{fig:phasediagram} we show the phase diagram divided in the different regions where a given parametrization applies.

While this procedure allows us to connect the solution of the eight-vertex model to each point of the full phase-diagram of the $XYZ$ chain, it treats each region separately and introduces artificial discontinuities at the boundaries between every two regions, as indicated by the {\it max} and {\it min} functions and the associated absolute values. In the next section we will show that, with a suitable analytical continuation of the elliptic parameters, we can extend the parametrization of a given region to the whole phase diagram. As we discussed, some discontinuities will remain unavoidable, but their locations will become a branch choice. Moreover, we will show that this extension numerically coincides with Baxter's prescription.

\section{Analytical extension of Baxter's solution}
\label{sect:newparameters}

To streamline the computation, it is convenient to perform a Landen transformation on (\ref{eq:XYZ3bis}) and to switch to new parameters $(u,l)$ instead of $(k,\lambda)$:
\be
   l \equiv {2 \sqrt{k} \over 1 + k} \; , \qquad \qquad
   u \equiv (1+k) \lambda \; ,
   \label{landen}
\ee
so that\footnote{See, for instance, table 8.152 of [\onlinecite{gradshteyn}]}
\bea
   \Gamma & = & {1 + k \; \sn^2 (\ii \lambda ; k) \over 1 - k \; \sn^2 (\ii \lambda ; k) }
   = {1 \over \dn ( \ii u; l ) } \; , 
   \label{Gnd} \\
   \Delta & = & - { \cn (\ii \lambda ; k) \; \dn (\ii \lambda ; k) \over
   1 - k \; \sn^2 (\ii \lambda ; k) }
   = - { \cn (\ii u ; l) \over \dn (\ii u ; l) } \; .
   \label{Dcd} 
\eea
The natural domain of $(\lambda,k)$ corresponds to 
\be
   0 \le u \le 2 K(l')  \qquad \qquad
   0 \le l \le 1 \; .
   \label{zlnatdomain}
\ee
We also have the following identities
\be
\label{eps1}
   \hskip -1.5cm l' = \sqrt{1 - l^2} = { 1 - k \over 1 +k} \; , \qquad
   K (l) = (1 +k) K (k) \; , \qquad
   K (l') = {1 + k \over 2} K (k') \; ,
\ee
from which it follows that the Landen transformation doubles the elliptic parameter $\tau(k) = \ii {K(k') \over K(k)} = 2 \ii {K(l') \over K(l)} = 2 \tau(l)$.

Using (\ref{ellid3}) we have  
\begin{equation}
    \label{ste8bis}
    l = \displaystyle{\sqrt{ {1-\Gamma^{2} \over \Delta^{2}-\Gamma^{2} } } } \; .
\end{equation}
Furthermore, elliptic functions can be inverted in elliptic integrals and from (\ref{Dcd}) we have (see Appendix A for the definition of the incomplete elliptical integral $F$):
\begin{equation}
    \label{ste8tris}
    \ii u = \int_{-1}^{-\Gamma} { \de t \over \sqrt{ (1 - t^2)( 1 - l^{\prime 2} t^2)} }
    = \ii \bigg[ K(l') - F (\arcsin \Gamma; l') \bigg]  \; .
\end{equation}
Together (\ref{ste8bis}) and (\ref{ste8tris}) invert (\ref{Gnd}) and (\ref{Dcd}) and, together with (\ref{st11}) and (\ref{st12}), give the value of Baxter's parameters for equivalent points of the phase diagram. It is important to notice that these identities ensure that the domain (\ref{zlnatdomain}) maps into $|\Gamma| \le 1$ and $\Delta \le -1$ and vice versa.

The idea of the analytical continuation is to use (\ref{ste8bis}) to extend the domain of $l$ beyond its natural regime. To this end, let us start from a given region, let say ${\rm I}_a$ in Fig. \ref{fig:phasediagram}, with $|\tJ_z| \le 1$ and $\tJ_y \ge 1$. By (\ref{st12}) we have:
\be
   \hskip -1.5cm 
   \tJ_y (u,l) = - \Delta = {\cn ( \ii u ; \,l) \over \dn ( \ii u ; \,l ) } \; , 
   \qquad \qquad
   \tJ_z (u,l) = -\Gamma = - {1 \over \dn ( \ii u ; \,l ) } \; ,
   \label{Iaparameter}
\ee
and thus
\begin{equation}
    \label{lzI}
   l = \sqrt{ {1 - \tJ_z^2 \over \tJ_y^2 - \tJ_z^2 } }  \; ,
   \qquad \qquad
    \ii u= \int_{-1}^{\tJ_z} { \de t \over \sqrt{ (1 - t^2)( 1 - l^{\prime 2} t^2)} } \; .
\end{equation}
Notice that within region ${\rm I}_a$ we can recast the second identity as
\be
   u = K (l')  + F (\arcsin \tJ_z; l') \; ,
\ee
which makes $u$ explicitly real and $0 \le u \le 2 K(l')$. 

We now take (\ref{lzI}) as the definition of $l$ and use it to extend it over the whole phase diagram.
For instance, in region ${\rm II}_a$ ($0 \le \tJ_y \le 1$, $\tJ_z^2 \le \tJ_y^2$), using (\ref{lzI}) we find $l >1$. To show that this analytical continuation gives the correct results, we write $l =1 / \tilde{l}$, so that $0 \le \tilde{l} \le 1$ and use (\ref{jacobiSTS}) in (\ref{Iaparameter}), which gives
\be
   \tJ_y (u,l) = {\dn (\ii \tilde{u} ; \,\tilde{l} ) \over \cn (\ii \tilde{u} ; \,\tilde{l}  ) }\;, 
   \qquad \qquad 
   \tJ_z (u, l) = -{1 \over \cn ( \ii \tilde{u} ;\tilde{l}  ) } \; , 
\ee
which reproduces Baxter's definitions in (\ref{st11}):  $\Gamma = -{\tJ_z \over \tJ_y}$, $\Delta = - {1 \over \tJ_y}$. Note that we also rescale the argument $\tilde{u} = l u$, so that $0 \le \tilde{u} \le 2 K(\tilde{l}')$. This shows that we can use (\ref{Iaparameter}) to cover both regions ${\rm I}_a$ and ${\rm II}_a$, by letting $0 \le l < \infty$ (note that $l=1$ corresponds to the boundary between the two regions). 

We can proceed similarly for the rest of the phase diagram. However, the second part of (\ref{lzI}) needs some adjustment.
Since elliptic integrals are multivalued functions of their parameters, to determine the proper branch of the integrand one starts in the principal region and follows the analytical continuation. The result of this procedure can be summarized as
\bea
   \ii u & = & \ii \bigg[ K(l') + F (\arcsin \tJ_z; l') \bigg] 
   + \bigg[ 1 - \sgn(\tJ_y) \bigg]  K (l) 
   \nonumber \\
   & = & F (\arcsin \tJ_y; l) - K(l) + \ii \bigg[ 1 + \sgn(\tJ_z) \bigg]  K (l')  \; ,
  \label{zparam}
\eea
where all the elliptic integrals here are taken at their principal value. The choice of either branch discontinuity corresponds to different ways of projecting the 3D projective space of parameters into the 2D phase space. We introduced this phenomenon in the previous section and we will come back to these jumps in a few paragraphs. Close to $\tJ_z \simeq 0$, the first line of (\ref{zparam}) is continuous, while the second has a jump. The opposite happens for $\tJ_y \simeq 0$, but, due to the periodicity properties of (\ref{Gnd}, \ref{Dcd}), both expressions are proper inversions of (\ref{Iaparameter}) valid everywhere. 

\begin{figure}
   \centering\includegraphics[width=\columnwidth]{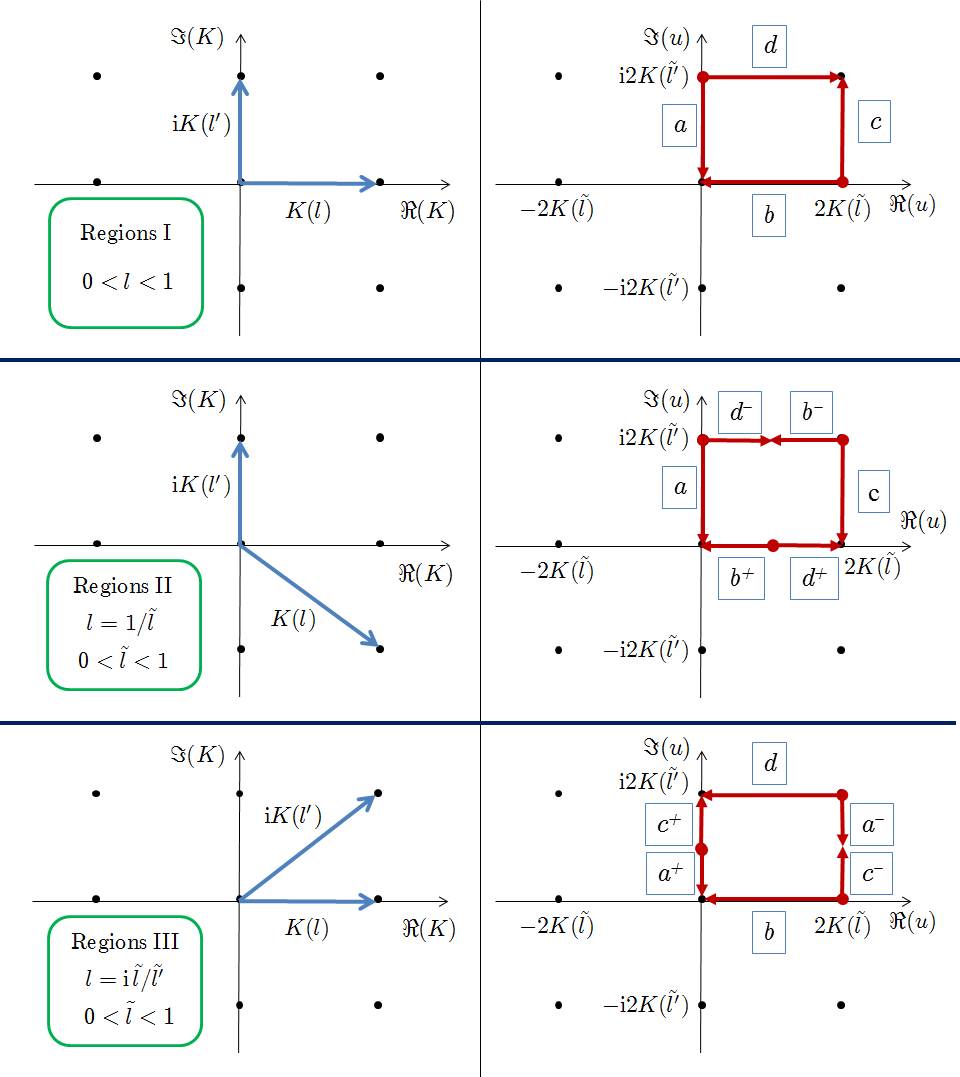}
   \caption{(Color online) A schematic representation of the analytical extension of Baxter's parametrization (\ref{zlparameter}) on the full phase diagram of Fig. \ref{fig:phasediagram}. Left: the directions of the quater-periods $K(l)$ and $\ii K(l')$ in ${\rm I}$, ${\rm II}$, and ${\rm III}$ regions. Right: the paths $a$, $b$, $c$, and $d$ of the argument of the elliptic functions. The argument always runs along the sides of the same rectangle, but, due to the branch discontinuities introduced in representing a three-dimensional space with two parameters, the path looks different in the three cases. The corners of the rectangle toward which the arrows point represent the tricritical ferromagnetic points, while the corners with the circles from which arrows start are the tricritical conformal points.}
   \label{fig:paths}
\end{figure}

\begin{figure}
   \dimen0=\textwidth
   \advance\dimen0 by -\columnsep
   \divide\dimen0 by 3
   \noindent\begin{minipage}[t]{\dimen0}
   {\centering \qquad \quad Regions ${\rm I} \quad \qquad$} 
   \includegraphics[width=\columnwidth]{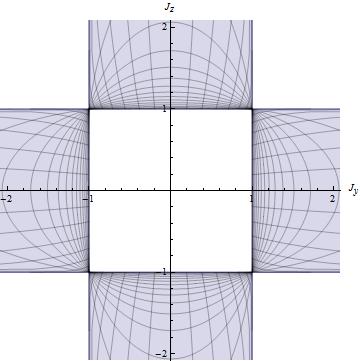}
   \end{minipage}
   \hfill
   \begin{minipage}[t]{\dimen0}
   {\centering \qquad \quad Regions ${\rm II} \quad \qquad$} 
   \includegraphics[width=\columnwidth]{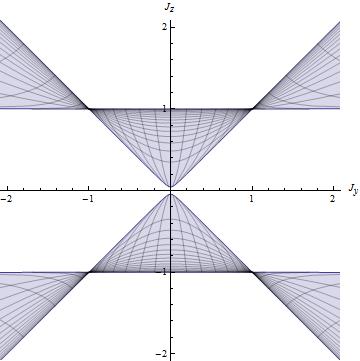}
   \end{minipage}
   \hfill
   \begin{minipage}[t]{\dimen0}
   {\centering \qquad \quad Regions ${\rm III} \quad \qquad$} 
   \includegraphics[width=\columnwidth]{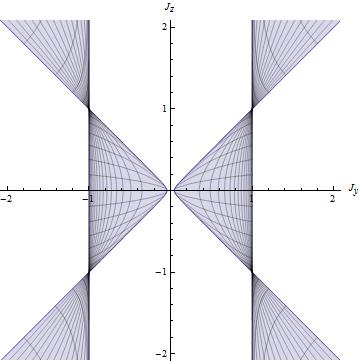}
   \end{minipage}
   \begin{minipage}[t]{\textwidth}
\caption{(Color online) Contour plots of the $(u,l)$ lines in the different regions of the phase diagram, as given by (\ref{zlparameter}).}
   \label{fig:contours}
   \end{minipage}
\end{figure}

Thus, we accomplished to invert (\ref{Iaparameter}) and to assign a pair of $(u,l)$ to each point of the phase diagram $(\tJ_y, \tJ_z)$, modulo the periodicity in $u$ space. The analysis of these mappings shows an interesting structure. From (\ref{lzI}) it follows that that regions ${\rm I}$'s have $0 \le l \le 1$, while ${\rm II}$'s have $l \ge 1$, which can be written as $l = 1 / \tilde{l}$, with $0 \le \tilde{l} \le 1$. Finally, the regions ${\rm III}$'s have purely imaginary $l = \ii \, \tilde{l}/\tilde{l'}$, with $0\le \tilde{l} \le 1$. In each region, the argument runs along one of the sides of a rectangle in the complex plane of sides $2 K(\tilde{l})$ and $\ii 2 K(\tilde{l}')$. Thus, we can write
\be
   \hskip -1.5cm 
   \tJ_y (u,l) =  {\cn ( \zeta \: u(z) ; \,l) \over \dn ( \zeta \: u(z) ; \,l ) } \; , \qquad 
   \tJ_z (u,l) =  {1 \over \dn ( \zeta \: u (z) ; \,l ) } \; ,
   \qquad  0 \le z \le \pi\; ,
   \label{zlparameter}
\ee
where $\zeta =1, \tilde{l}, \tilde{l}'$ in regions ${\rm I}$'s,  ${\rm II}$'s, and ${\rm III}$'s respectively,  and with $z$ defined implicitly as follows.  The path of $u(z)$ runs along the sides of a rectangle drawn on the half periods $2 K(\tilde{l})$ and $2 \ii K(\tilde{l}')$, as shown in Fig. \ref{fig:paths}. For instance, for type ${\rm I}$ we have $u (z) = \ii (\pi - z) {2 \over \pi} K(\tilde{l}')$ for regions of type $a$;  $u (z) = (\pi - z) {2 \over \pi} K(\tilde{l})$ for type $b$'s; $u (z) =2 K(\tilde{l}) +  \ii z {2 \over \pi} K(\tilde{l}')$ for regions of type $c$; and $u (z) = \ii 2 K(\tilde{l}') + z {2 \over \pi} K(\tilde{l})$ for regions of type $d$. Using the same paths for ${\rm II}$'s and ${\rm III}$'s would reverse the role of the ferro-/antiferromagnetic nature of the phases and thus we are forced to introduce jump discontinuities in the paths, see Fig. \ref{fig:paths}; these are the aforementioned singularities introduced by the 2D representation of the three-dimensional projective space of parameters. In the next section, we will work directly on the latter and we will show that the modular group naturally generates the regular paths of regions ${\rm I}$ for all the cases. In Fig. \ref{fig:contours} we draw the contour lines of $(u,l)$ in the $(\tJ_y, \tJ_z)$ plane, as given by (\ref{zlparameter}).

The rectangular shape for the path of the argument is not surprising, since it is known that only purely real or purely imaginary arguments (modulo a half periodicity) ensure the reality of an elliptic function (and thus that the coupling constants are real and the Hamiltonian Hermitian).
The validity of these analytical continuations can be checked, like we just did for region ${\rm II}_a$, using the relations connecting elliptic functions of different elliptic parameters (see Appendix \ref{app:elliptic} or [\onlinecite{lawden}]). All these relations are rooted in the modular invariance of the torus on which elliptic functions are defined. In the next section, we will show that the action of the modular group in covering the phase diagram of the $XYZ$ model mirrors the symmetry of the model.

Before we proceed, we remark that, while it is possible that the validity of Baxter parametrization outside of its natural domain has been observed before by other authors' analysis, we did not find any reference to it in the literature. We show in detail how to construct the complete extension to the whole phase diagram, which is condensed in Table \ref{transformationtable} in the next section.

\section{Action of the modular group}
\label{sec:modular}

In the previous section, we showed that Baxter's parametrization of the parameters of the $XYZ$ chain can be analytically extended outside of the principal regime to cover the whole phase diagram and thus that any path in the real two-dimensional $(\tJ_y, \tJ_z)$ space corresponds to a path of $(u,l)$ in ${\mathbb{C}}^2$. This extension is due to the fact that the analytical continuation of elliptic functions in the complex plane can be related to the action of the modular group. We refer to Ref. \onlinecite{lawden} for a detailed explanation of the relation between elliptic functions and the modular group and to Appendix \ref{app:elliptic} for a collection of useful identities. 

Points related by a modular transformation correspond to the same $(\Gamma,\Delta)$ point in the mapping to the eight-vertex model (\ref{st11}) and (\ref{st12}). We now want to show that the modular group can be connected to the symmetries of the model that we discussed in the Introduction. Let us start again in region ${\rm I_a}$, but to make the modular structure more apparent, this time let us write the parametrization in terms of $\theta$ functions using expression (\ref{jacobidef}):
\begin{subequations}
\bea
  \tJ_{y} & = & {J_y \over J_x} = \frac{\theta_{3}(0|\tau)}{\theta_{2}(0|\tau)}
  \frac{\theta_{2} \left[ (z - \pi) \tau|\tau \right]}{\theta_{3} \left[ (z - \pi) \tau|\tau \right]} \; , \\
  \tJ_{z} & = & {J_z \over J_x} = \frac{\theta_{3}(0|\tau)}{\theta_{4}(0|\tau)}
  \frac{\theta_{4} \left[ (z - \pi) \tau|\tau \right]}{\theta_{3} \left[ (z - \pi) \tau|\tau \right]} \; ,
\eea
\end{subequations}
where $0 \le z \le \pi$, and we used (\ref{thetaperiod}).

In the previous section, we noticed that using the homogeneous coordinates $(\tJ_z, \tJ_y)$ introduces artificial discontinuities in the phase diagram. Thus, we now revert back to the original three-dimensional projective space of parameters and write
\begin{subequations}
\bea
    J_x & = & J \:
    \frac{\theta_{3} \left[ (z - \pi) \tau|\tau \right]}{\theta_{3} \left( 0 |\tau \right)} \; , \\
    J_y & = & J \:
    \frac{\theta_{2} \left[ (z - \pi) \tau|\tau \right]}{\theta_{2} \left( 0 |\tau \right)} \; , \\
    J_z & = & J \:
    \frac{\theta_{4} \left[ (z - \pi) \tau|\tau \right]}{\theta_{4} \left( 0 |\tau \right)} \; ,
\eea
\label{jthetas}
\end{subequations}
where $J$ is an additional normalization which we will need.

As the Jacobi elliptic functions are doubly-periodic, the torus on which they are defined can be viewed as the quotient $\mathbb{C}/\mathbb{Z}^2$. In region ${\rm I_a}$, for $0 \le l \le 1$, the fundamental domain is the rectangle with sides on the real and imaginary axis, of length respectively $\omega_1 = 4 K(l)$ and $\omega_3 = \ii 4K(l')$, and the path followed by $z$ is a half period along the imaginary axis. A modular transformation turns the rectangle into a parallelogram constructed on the same two-dimensional lattice, but leaves the path of the argument untouched. To this end, we extend the representation of the modular group to keep track of the original periods and introduce the two vectors $\sigma_R$ and $\sigma_I$, which points along the original real and imaginary axis. In ${\rm I_a}$, these two vectors are $\sigma_R = {\omega_1 \over \omega_1} = 1$ and $\sigma_I = {\omega_3 \over \omega_1} = \tau$ (we remind that, when using $\theta$ functions, every length is normalized by the periodicity on the real axis, i.e., $\omega_1$). We will also need to extend the modular group and thus we introduce the additional complex number $\phi = -\pi \sigma_I$, which is a sort of an initial phase in the path of the argument. 

With these definitions, we write (\ref{jthetas}) as
\be
  \hskip-1cm
  J_x= J \: 
  \frac{\theta_{3} \left[ z \, \sigma_I + \phi |\tau \right]}{\theta_{3} \left( 0 |\tau \right)} \; , 
  \quad
  J_y= J \:  
  \frac{\theta_{2} \left[ z \, \sigma_I + \phi |\tau \right]}{\theta_{2} \left( 0  |\tau \right)} \; , 
  \quad
  J_z= J \: 
  \frac{\theta_{4} \left[ z \, \sigma_I + \phi |\tau \right]}{\theta_{4} \left( 0 |\tau \right)} \; .
\label{thetaJs}
\ee

The two generators of the modular group act on this extended representation as
\be
    {\bf T} \left( \begin{array}{c} \tau \cr \sigma_R \cr \sigma_I \cr \phi \end{array} \right) =
    \left( \begin{array}{c} \tau + 1\cr \sigma_R \cr \sigma_I \cr \phi \end{array} \right) \; , \qquad \qquad
    {\bf S} \left( \begin{array}{c} \tau \cr \sigma_R \cr \sigma_I \cr \phi \end{array} \right) =
    \left( \begin{array}{c} - {1 \over \tau} \cr - {\sigma_R \over \tau} \cr - {\sigma_I \over \tau} \cr - {\phi \over \tau} \end{array} \right) \; .
    \label{STgendef}
\ee
A general modular transformation which gives $\tau' = {c + d \, \tau \over a + b \, \tau}$ also does 
\be
  \hskip -1.5cm \sigma_R \to {1 \over a + b \, \tau} = d - b \, \tau' \, , \qquad 
  \sigma_I \to {\tau \over a + b\, \tau} = -c + a \, \tau' \, , \qquad 
  \phi \to {\phi \over a + b \, \tau} \, ,
\ee
i.e. it expresses the original real and imaginary periods in terms of the two new periods.

To see the action of these transformations in the space of parameters, let us start again from the subregion ${\rm I_a}$. Applying the ${\bf S}$ transformation ${\bf S} \tau = \tau_S = -{1 \over \tau}$, we have
\begin{subequations}
\bea
   {\bf S} J_x & = & \left( {\bf S} J \right)  
   \frac{\theta_{3} \left[ z - \pi |\tau_S \right]}{\theta_{3} \left( 0 |\tau_S \right)} 
   = \left( {\bf S} J \right) \eu^{{\ii \tau \over \pi} (z - \pi)^2}
   \frac{\theta_{3} \left[ (z - \pi) \tau |\tau \right]}{\theta_{3} \left( 0 |\tau \right)} 
   = J_x \; , 
   \\
  {\bf S} J_y & = & \left( {\bf S} J \right) 
  \frac{\theta_{2} \left[ z - \pi |\tau_S \right]}{\theta_{2} \left( 0 |\tau_S \right)} 
  = \left( {\bf S} J \right) \eu^{{\ii \tau \over \pi} (z - \pi)^2}
  \frac{\theta_{4} \left[ (z - \pi) \tau |\tau \right]}{\theta_{4} \left( 0 |\tau \right)} 
   =  J_z \; , 
  \\
  {\bf S} J_z & = & \left( {\bf S} J \right)  
  \frac{\theta_{4} \left[ z - \pi |\tau_S \right]}{\theta_{4} \left( 0 |\tau_S \right)} 
  = \left( {\bf S} J \right) \eu^{{\ii \tau \over \pi} (z - \pi)^2}
  \frac{\theta_{2} \left[ (z - \pi) \tau |\tau \right]}{\theta_{2} \left( 0 |\tau \right)} 
   = J_y \; , 
\eea
\label{SAction}
\end{subequations}
where we defined
\be
    {\bf S} J \equiv \eu^{-{\ii \tau \over \pi} (z - \pi)^2} \: J \; .
\ee
Equations (\ref{SAction}) show that $ {\bf S}$ corresponds to the exchange  $E_x$.

The other generator of the modular group ${\bf T} \tau = \tau_T = \tau +1$ gives
\begin{subequations}
\bea
   {\bf T} J_x & = &  \left( {\bf T} J \right) 
   \frac{\theta_{3} \left[ (z - \pi) \tau |\tau_T \right]}{\theta_{3} \left( 0 \tau |\tau_T \right)} 
   =  \left( {\bf T} J \right) 
   \frac{\theta_{4} \left[ (z - \pi) \tau | \tau \right]}{\theta_{4} \left( 0  \tau | \tau \right)}  
  = J_z  \, ,\\ 
   {\bf T} J_y & = &  \left( {\bf T} J \right) 
  \frac{\theta_{2} \left[ (z - \pi) \tau |\tau_T \right]}{\theta_{2} \left( 0 |\tau_T \right)} 
   =  \left( {\bf T} J \right) 
  \frac{\theta_{2} \left[ (z - \pi) \tau | \tau \right]}{\theta_{2} \left( 0 | \tau \right)}  
  = J_y \, ,\\ 
  {\bf T} J_z & = & \left( {\bf T} J \right) 
  \frac{\theta_{4} \left[ ( z - \pi ) \tau |\tau_T \right]}{\theta_{4} \left( 0 |\tau_T \right)} 
  = \left( {\bf T} J \right) 
  \frac{\theta_{3} \left[  (z - \pi ) \tau | \tau \right]}{\theta_{3} \left( 0 |\tau \right)}   
  = J_x \; ,
\eea
\label{TAction}
\end{subequations}
which shows that ${\bf T}$ corresponds to $ E_y$ and sets
\be
   {\bf T} J \equiv J \; .
\ee

The normalization constant $J$ is inconsequential, since it just sets the energy scale. However, to make the link between the generator of the modular group and the symmetry of the model more apparent and to maintain reality of the couplings, we choose it such that it cancels the phases of the ${\bf S}$ transformation. Such choice can be accommodated by setting, for instance,
\be
   J = J(z, \tau) \propto \Big( \theta_{2} \left[ (z - \pi) \tau |\tau \right]
  \theta_{3} \left[ (z - \pi) \tau |\tau \right]  \theta_{4} \left[ (z - \pi) \tau |\tau \right] \Big)^{-1/3} \; .
\ee

Finally, we introduce an additional operator ${\bf P}$:
\be
    {\bf P} \left( \begin{array}{c} \tau \cr \sigma_R \cr \sigma_I \cr \phi \end{array} \right) =
    \left( \begin{array}{c} \tau \cr \sigma_R \cr \sigma_I \cr \phi + \pi \left[ \sigma_R + \sigma_I \right] \end{array} \right) \; ,
    \label{Pdef}
\ee
which shifts by half a period in both directions the origin of the path of the argument. Its action is given by [see (\ref{thetaperiod})]
\begin{subequations}
\bea
   {\bf P} J_x & = &  \left( {\bf P} J \right) 
   \frac{\theta_{3} \left[ z \tau + \pi |\tau \right]}{\theta_{3} \left( 0 |\tau \right)} 
  = J_x  \, ,\\ 
   {\bf P} J_y & = &  \left( {\bf P} J \right) 
  \frac{\theta_{2} \left[ z \tau + \pi |\tau \right]}{\theta_{2} \left( 0 |\tau \right)} 
  = - J_y  \, ,\\ 
  {\bf P} J_z & = &   \left( {\bf P} J \right) 
  \frac{\theta_{4} \left[ z \tau + \pi |\tau \right]}{\theta_{4} \left( 0 |\tau \right)} 
  = - J_z \; ,
\eea
\label{PAction}
\end{subequations}
so that ${\bf P}$ corresponds to the $\pi$-rotation generator $P_x$.

The three operation ${\bf T}$, ${\bf S}$ and ${\bf P}$, with:
\be
  \hskip -2cm ({\bf S} \cdot {\bf T})^{3}=\mathbb{I} \, , \qquad
  {\bf S}^2 = \mathbb{I} \, , \qquad
  {\bf P}^2 = \mathbb{I} \, , \qquad
  {\bf P} \cdot {\bf S} = {\bf S} \cdot {\bf P} \, , \qquad
  {\bf P} \cdot {\bf T} = {\bf T} \cdot {\bf P} \, .
\ee
generate a $\mathbb Z _2$ extension of the modular group that can be identified \cite{Jones} with $PGL(2,\mathbb Z)$. We collect in Table \ref{transformationtable} the action of the different transformations generated by ${\bf T}$, ${\bf S}$ and ${\bf P}$ for the regions in the positive half plane, the other being given just by acting with the $\pi$ rotation.

\begin{table}
\begin{center}
\begin{tabular}{| c | c | c || c | c | c | c | c  || c | c | c |}
\hline
\multirow{2}{*}{\scriptsize{Region}} & {\scriptsize{Modular}}  & \multirow{2}{*}{$\phi$} & {\scriptsize{Couplings}} & \multirow{2}{*}{$\sigma_R$} & \multirow{2}{*}{$\sigma_I$}& \multirow{2}{*}{$\tau$} & \multirow{2}{*}{$l$}  & \multirow{2}{*}{$\phi$} & {\scriptsize{Modular}} &\multirow{2}{*}{\scriptsize{Region}} \cr
& {\scriptsize{Generator}} &  & &  & &  & & & {\scriptsize{Generator}} & \cr
\hline
${\rm I}_a$  & ${\bf I}$ & $-\pi \tau$ & $\{J_x, J_y, J_z\}$ & $1$ & $\tau$ & $\tau$ & $\tilde{l}$ 
& $\pi$ & {\bf P} & ${\rm I}_c$ \cr
\hline
${\rm I}_b$ & {\bf S} & $-\pi $ & $\{J_x, J_z, J_y\}$ & ${1 \over \tau}$ & $1$ & $-{1 \over \tau}$ & $\tilde{l}'$ 
& ${\pi \over \tau} $ & {\bf SP} & ${\rm I}_d$  \cr
\hline
${\rm III}_a$ & {\bf T} & $-\pi \tau $ & $\{J_z, J_y, J_x\}$ & $1$ & $\tau$ & $ \tau + 1$ & $\ii {\tilde{l} \over \tilde{l}'}$ 
& $\pi$ & {\bf TP}  & ${\rm III}_c$ \cr
\hline
${\rm II}_b$ & {\bf TS} & $-\pi$ & $\{J_z, J_x, J_y\}$ & ${1 \over \tau}$ & $1$ & ${\tau -1 \over \tau}$ & ${1 \over \tilde{l}'}$ 
& ${\pi \over \tau} $ & {\bf TSP} & ${\rm II}_d$   \cr
\hline
${\rm III}_b$ & {\bf ST} & ${- \pi \tau \over \tau +1} $ & $\{J_y, J_z, J_x\}$ & ${1 \over \tau + 1}$ & ${\tau \over \tau +1}$ & ${-1 \over \tau+1}$ & $\ii {\tilde{l}' \over \tilde{l}}$  
& $ {\pi \over \tau +1} $ & {\bf STP} & ${\rm III}_d$   \cr
\hline
${\rm II}_a$ & {\bf TST} & ${- \pi \tau \over \tau +1} $ & $\{J_y, J_x, J_z\}$ & ${1 \over \tau + 1}$ & ${\tau \over \tau +1}$ & ${\tau \over \tau+1}$ & ${1 \over \tilde{l}}$ 
& $ {\pi \over \tau +1} $ & {\bf TSTP} & ${\rm II}_c$  \cr
& {\bf STS} & ${- \pi \tau \over \tau -1} $ & & ${1 \over \tau - 1}$ & ${\tau \over \tau -1}$ & ${\tau \over \tau-1}$ & ${1 \over \tilde{l}}$  
& $ {\pi \over \tau - 1} $ & {\bf STSP} & \cr
\hline
\end{tabular}
\caption{List of the action of modular transformations, according to (\ref{STgendef}) and (\ref{Pdef}), in mapping the phase diagram, starting from ${\rm I}_a$ with the parametrization given by (\ref{thetaJs}).}
\label{transformationtable}
\end{center}
\end{table}

A few additional comments are in order. In the previous section we gave a prescription for the path of the argument around a rectangle; see (\ref{zlparameter}) and Fig. \ref{fig:contours}. This structure is explained by Table \ref{transformationtable}: the ${\bf S}$ generator interchanges the role of the axes and thus the path that runs along the imaginary axis in the new basis runs along the real one, consistently with the prescription given for regions of type $b$ and $d$. The action of ${\bf P}$ is to shift the path and thus interchanges regions of type $a$ with $c$ and $b$ with $d$ (and viceversa). The role of ${\bf P}$ can be thought of as that of promoting the modular group to its discrete affine.

In conclusions, we see that the whole line $l^2 \in (- \infty, \infty)$ can be used in (\ref{zlparameter}), which divides in three parts, one for each of the regions of type ${\rm I}$, ${\rm II}$, or ${\rm III}$. It is known, that, for each point on this line, only if the argument of the elliptic functions is purely imaginary or purely real (modulo a half periodicity)  are we guaranteed that the parameters of the model are real. Hence the two (four) paths, related by ${\bf S}$ (and ${\bf P}$) duality. The paths of the argument are compact, due to periodicity. Thus, the mapping between $(u,l)$ and $(J_y, J_z)$ is topologically equivalent to $\mathbb{R}^2 \to \mathbb{R} \otimes \mathbb{C} \otimes \mathbb{Z}_2 \otimes \mathbb{Z}_2$. Extending the mapping by considering a generic complex argument $u$ would yield a non-Hermitian Hamiltonian, but would preserve the structure that makes it integrable. We do not have a physical interpretation for such a nonunitary extension, but its mathematical properties indicates that it might be worth exploring this possibility.

\section{Conclusions and outlook}
\label{sec:conclusions}

The possibility to write an exact solution for the $XYZ$ spin chain is based on its relation with the eight-vertex model. However, the relation between the parameters $(\Gamma, \Delta)$ of the latter is not one to one with the couplings $(J_x, J_y, J_z)$ of the former (and with the Boltzmann weights of the classical model as well): indeed such mapping changes in the different regions of the phase diagram; see Eqs. (\ref{st11}) and (\ref{st12}). In this paper, we have shown that this re-arrangement procedure can be  reproduced by means of the {\it natural} analytical extension of the solution valid in a given initial region. We have also given a prescription that relates the parameter $(u,l)$ of the elliptic functions and the physical parameter of the $XYZ$ model in the whole of the phase diagram and provided an inversion formula valid everywhere.

Clearly symmetries in the space of parameters become manifest as invariance properties of observables and correlations. For example, in Ref. \onlinecite{Ercolessi2011} we have analytically calculated entanglement entropies for the $XYZ$ model and in particular obtained an expression for  the corrections to the leading terms of  R\'enyi entropies that emerge as soon as one moves into the massive region, starting from the critical (conformal) line. Also, by making explicit use of such a kind of parametrization, we have shown that finite-size and finite-mass effects give rise to different contributions (with different exponents), thus violating simple scaling arguments. This topic deserves further analytical and numerical studies that we plan to present in a future paper. 

The main novelty of the approach presented in this paper lies in the connection between analytic continuation in the parameter space and the action of the modular group. This connection allowed us to show that a certain  extension of the modular group realizes in parameter space the physical symmetries of the model. This symmetries are in the form of {\it dualities}, since they connect points with different couplings, and thus could be exploited to compute different correlation functions, whose operator content is related by these dualities, potentially simplifying the operation. 

Modular invariance plays a central role in the structure and integrability of Conformal Field Theories in $1 + 1$ dimension. On its critical lines, the $XYZ$ model is described by a $c=1$ CFT and thus, in the scaling limit, its partition function is modular invariant in real space. Moreover, it is known \cite{difrancesco} that modular invariance plays a central role in organizing the operator content of a CFT. Our results show that, even in the gapped phase, the $XYZ$ chain is also a modular invariant in parameter space, due to the symmetries of the model. Elliptic structures are common in the solution of integrable models and it is tempting to speculate that their modular properties do encode in general their symmetries and thus the class of integrability-preserving relevant perturbations that drive the system away from criticality. In fact, this observation might help in interpreting the long-standing mystery for which the partition function of integrable models in their gapped phases, can be organized in characters of Virasoro representations, with the finite-size/-temperature scale set, instead, by the mass parameter. \cite{JimboMiwa, BauerSaleur, cardy1988} This is a peculiar observation that has so far eluded fundamental explanation: our work hints that the symmetries of an integrable system can manifest themselves with a modular structure in parameter space and this would constrain the form of the partition function in the same way as modular invariance in coordinate space operates to organize the spectrum of critical theories. While some recent results from the study of entanglement entropies already support this line of thinking, \cite{ercolessi2012, deluca2013} we are currently exploring to which extent this speculation can hold true in general.

\section*{Acknowledgments}

F.F. thanks P. Glorioso for interesting and fruitful discussions. This work was supported in part by two INFN COM4 grants (FI11 and NA41). F.F. was supported by a Marie Curie International Outgoing Fellowship within the 7th European Community Framework Programme (FP7/2007-2013) under Grant No. PIOF-PHY-276093.

\appendix

\section{Elliptic Functions}
\label{app:elliptic}

Elliptic functions are the extensions of the trigonometric functions to work with doubly periodic expressions, i.e., such that
\be
   f \left( z + 2 n \omega_1 + 2 m \omega_3 \right) = f \left( z + 2 n \omega_1 \right) = f (z) \; ,
\ee
where $\omega_1, \omega_3$ are two different complex numbers and $n,m$ are integers (following Ref. \onlinecite{lawden}, we have
$ \omega_1 + \omega_2 + \omega_3 =0$).

The Jacobi elliptic functions are usually defined through the pseudoperiodic {\it $\theta$ functions}:
\begin{subequations}
\bea
 \theta_1(z;q) &=& 2 \sum_{n = 0}^\infty (-1)^n q^{(n + 1/2)^2} \sin [(2n + 1) z] \; ,\\ \label{theta1}
 \theta_2(z;q) &=& 2 \sum_{n = 0}^\infty q^{(n + 1/2)^2} \cos[(2n+1) z] \; ,\\
 \theta_3(z;q) &=&  1 + 2 \sum_{n=1}^\infty q^{n^2} \cos (2 n z) \; , \\
 \theta_4(z;q) &= & 1 + 2 \sum_{n=1}^\infty (-1)^n q^{n^2} \cos (2 n z) \; ,
\eea
\label{thetadef}
\end{subequations}
where the {\it elliptic nome} $q$ is usually written as $q \equiv \eu^{\ii \pi \tau}$ in terms of the elliptic parameter $\tau$ and accordingly
\be
  \theta_j (z | \tau) \equiv \theta_j (z;q) \; .
\ee
The periodicity properties of the $\theta$ functions are
\begin{subequations}
\bea
 \hskip -1cm \theta_1(z | \tau) &=&  - \theta_1 (z +  \pi | \tau) = - \lambda \theta_1 (z + \pi \tau | \tau) =
 \: \: \: \lambda \theta_1 (z + \pi + \pi \tau | \tau ) \; ,\\ 
 \hskip -1cm \theta_2(z | \tau) &=&  - \theta_2 (z +  \pi | \tau) = \: \: \: \lambda \theta_2 (z + \pi \tau | \tau) =
 - \lambda \theta_2 (z + \pi + \pi \tau | \tau) \; ,\\ 
 \hskip -1cm \theta_3(z | \tau) &= & \: \: \: \theta_3 (z +  \pi | \tau) = \: \: \: \lambda \theta_3 (z + \pi \tau | \tau) =
 \: \: \: \: \lambda \theta_3 (z + \pi + \pi \tau | \tau) \; ,\\ 
\hskip -1cm  \theta_4(z | \tau) &=&  \: \: \: \theta_4 (z +  \pi | \tau) = - \lambda \theta_4 (z + \pi \tau | \tau) =
 - \lambda \theta_4 (z + \pi + \pi \tau | \tau) \; ,
\eea
\label{thetaperiod}
\end{subequations}
where $\lambda \equiv q \eu^{2 \ii z}$.

Moreover, incrementation of $z$ by the half periods ${1 \over 2} \pi, {1 \over 2} \pi \tau$, and ${1 \over 2} \pi (1 + \tau)$ leads to 
\begin{subequations}
\bea
 \hskip -1.5cm \theta_1(z | \tau) &=&  - \theta_2 (z +  {\textstyle{1 \over 2}} \pi | \tau) = - \ii \mu \theta_4 (z + {\textstyle{1 \over 2}} \pi \tau | \tau) =
 -\ii \mu \theta_3 (z + {\textstyle{1 \over 2} \pi + {1 \over 2}}  \pi \tau | \tau ) \; ,\\ 
 \hskip -1.5cm \theta_2(z | \tau) &=&  \: \: \:  \theta_1 (z +  {\textstyle{1 \over 2}} \pi | \tau) = \: \: \: \: \mu \theta_3 (z + {\textstyle{1 \over 2}} \pi \tau | \tau) =
 \: \: \: \: \mu \theta_4 (z + {\textstyle{1 \over 2} \pi + {1 \over 2}}  \pi \tau | \tau ) \; ,\\ 
 \hskip -1.5cm \theta_3(z | \tau) &=&  \: \: \:  \theta_4 (z +  {\textstyle{1 \over 2}} \pi | \tau) = \: \: \:  \: \mu \theta_2 (z + {\textstyle{1 \over 2}} \pi \tau | \tau) =
 \: \: \: \: \mu \theta_1 (z + {\textstyle{1 \over 2} \pi + {1 \over 2}}  \pi \tau | \tau ) \; ,\\ 
 \hskip -1.5cm \theta_4(z | \tau) &=&  \: \: \: \theta_3 (z +  {\textstyle{1 \over 2}} \pi | \tau) = - \ii \mu \theta_1 (z + {\textstyle{1 \over 2}} \pi \tau | \tau) =
 \: \: \:  \ii \mu \theta_2 (z + {\textstyle{1 \over 2} \pi + {1 \over 2}}  \pi \tau | \tau ) \; ,
\eea
\label{thetaphalfperiod}
\end{subequations}
with $\mu \equiv q^{1/4} \eu^{\ii z}$.

The Jacobi elliptic functions are then defined as
\begin{subequations}
\bea
 \sn (u; k) & = &  {\theta_3 (0 | \tau) \over \theta_2 (0 | \tau) } \; {\theta_1 (z | \tau) \over \theta_4 (z | \tau) } \; ,\\ 
 \cn (u; k) & =  & {\theta_4 (0 | \tau) \over \theta_2 (0 | \tau) } \; {\theta_2 (z | \tau) \over \theta_4 (z | \tau) } \; ,\\ 
 \dn (u; k) & =  & {\theta_4 (0 | \tau) \over \theta_3 (0 | \tau) } \; {\theta_3 (z | \tau) \over \theta_4 (z | \tau) } \; ,
\eea
\label{jacobidef}
\end{subequations}
where $u = \theta_3^2 (0 | \tau) \, z = {2 \over \pi} \, K (k) \, z$, $\tau \equiv \ii {K (k') \over K (k)} $, $k' \equiv \sqrt{ 1 -k^2}$, and 
\be
   K (k) \equiv \int_0^1 { \de t \over \sqrt{ (1 - t^2) (1-k^2 t^2)}}
   = \int_0^{\pi \over 2} { \de \theta \over \sqrt{1 - k^2 \sin^2 \theta}}
   \label{Kdef}
\ee
is the elliptic integral of the first type. 
Additional elliptic functions can be generated with the following two rules: denote with $p,q$ the letters $s, n, c, d$, then
\be
  {\rm pq} (u;k) \equiv {1 \over {\rm qp} (u;k) } \; , \qquad \qquad
  {\rm pq} (u;k) \equiv {{\rm pn} (u;k) \over {\rm qn} (u;k)} \; .
\ee

For $0 \le k \le 1$, $K(k)$ is real, $\ii K^\prime (k) \equiv \ii K (k')$ is purely imaginary and together they are called 
{\it quarter periods}. The Jacobi elliptic functions inherit their periodic properties from the $\theta$ functions, thus:
\begin{subequations}
\bea
 \hskip-2cm \sn (u;k) &=&  - \sn (u + 2K;k) = \: \: \: \sn (u + 2 \ii K^\prime;k) = - \sn (u +2K +2\ii K^\prime; k ) \; ,\\ 
 \hskip-2cm \cn (u;k) &=&  - \cn (u + 2K;k) = - \cn (u + 2 \ii K^\prime;k) = \: \: \: \cn (u +2K +2\ii K^\prime; k ) \; ,\\ 
 \hskip-2cm \dn (u;k) &=&  \: \: \: \dn (u + 2K;k) = - \dn (u + 2 \ii K^\prime;k) = - \dn (u +2K +2\ii K^\prime; k ) \; .
\eea
\label{jacobiperiod}
\end{subequations}

The inverse of an elliptic function is an elliptic integral. The {\it incomplete elliptic integral of the first type} is usually written as
\be
   F (\phi;k) = \int_0^\phi {\de \theta \over \sqrt{1 - k^2 \sin^2 \theta} } 
   = \sn^{-1} \left( \sin \phi ; k \right) \; ,
   \label{Fdef}
\ee
and it is the inverse of the elliptic $\sn$. Clearly, $F \left( {\pi \over 2}; k \right) = K (k)$. Further inversion formulas for the elliptic functions can be found in Ref. \onlinecite{lawden}.

Additional important identities include
\be
   k = k ( \tau ) =  {\theta_2^2 (0 | \tau) \over \theta_3^2 (0 | \tau) } \; , \qquad \qquad
   k' = k' ( \tau ) =  {\theta_4^2 (0 | \tau) \over \theta_3^2 (0 | \tau) } \; , 
\ee
and
\bea
    \sn^2 (u;k) + \cn^2 (u;k) & = & 1 \; , \label{ellid1}  \\
    \dn^2 (u;k) + k^2 \sn^2 (u;k) & = & 1 \; , \label{ellid2}  \\
    \dn^2 (u;k) - k^2 \cn^2 (u;k) & = & k^{\prime 2} \label{ellid3} \; .
\eea    

Taken together, the elliptic functions have common periods $2 \omega_1 = 4 K(k)$ and $2\omega_3 =  4 \ii K^\prime (k)$. 
These periods draw a lattice in $\mathbb{C}$, which has the topology of a torus. While the natural domain is given 
by the rectangle with corners $(0,0),(4K,0),(4K,4 \ii K^\prime), (0,4 \ii K^\prime)$,  any other choice would be exactly equivalent. 
Thus, the parallelogram defined by the half periods
\be
   \omega_1^\prime = a \, \omega_1 + b \, \omega_3 \; , \qquad \quad
   \omega_3^\prime = c \, \omega_1 + d \, \omega_3 \; ,
\ee
with $a, b, c, d$ integers such that $ad - bc = 1$, can suit as a fundamental domain. This transformation, which also changes the 
elliptic parameter as
\be
   \tau^\prime = {\omega_3^\prime \over \omega_1^\prime } = { c + d \, \tau \over a + b \, \tau } \; ,
\ee
can be cast in a matrix form as
\be
   \left( \begin{array}{c} \omega_1^\prime \\ \omega_3^\prime \end{array} \right) =
   \left( \begin{array}{c c} a & b \\ c & d \end{array} \right)
   \left( \begin{array}{c} \omega_1 \\ \omega_3 \end{array} \right)
\ee
and defines a {\it modular transformation}. The modular transformations generate the {\it modular group} ${\rm PSL}(2,\mathbb{Z})$.
Each element of this group can be represented (in a {\bf not} unique way) by a combination of the two transformations
\bea
   {\bf S} = \left( \begin{array}{c c} 0 & 1 \\\ -1 & 0 \end{array} \right) & \quad \longrightarrow \quad & \tau^\prime = - {1 \over \tau} \; , \\
   {\bf T} = \left( \begin{array}{c c} 1 & 0 \\\ 1 & 1 \end{array} \right) & \quad \longrightarrow \quad & \tau^\prime =  \tau + 1 \; ,
\eea
which satisfy the defining relations
\be
    {\bf S}^2 = {\bf 1} \; , \qquad \qquad \left( {\bf ST} \right)^3 = {\bf 1} \; .
\ee

The transformation properties for the ${\bf S}$ transformation are
\begin{subequations}
\bea
 \theta_1 \left( z | - {1 \over \tau} \right) &=& -\ii(\ii \tau)^{\frac 1 2} \eu^{\frac{\ii \tau z^2}{\pi}}
 \theta_1( \tau z | \tau) \; , \\
 \theta_2 \left( z |  - {1 \over \tau} \right) &=& (-\ii \tau)^{\frac 1 2} \eu^{\ii \tau z^2 \over \pi}
 \theta_4 (\tau z | \tau) \; , \\
 \theta_3 \left( z  |  - {1 \over \tau} \right) &=& (-\ii \tau)^{\frac 1 2} \eu^{\ii \tau z^2 \over \pi}
 \theta_3 (\tau z | \tau) \; , \\
 \theta_4 \left (z |  - {1 \over \tau} \right) &=& (-\ii \tau)^{\frac 1 2} \eu^{\ii \tau z^2 \over \pi} \theta_2
(\tau z | \tau) \; ,
\eea
\label{SthetaTrans}
\end{subequations}
from which follows
\be
   k \left( - {1 \over \tau} \right) = k^\prime ( \tau ) \; , \qquad \qquad \qquad
   k^\prime  \left( - {1 \over \tau} \right) = k ( \tau ) \; ,
\ee
and
\be
  K \left( - {1 \over \tau} \right) = K' ( \tau) = K (k') \; , \qquad 
  \ii K^\prime \left( - {1 \over \tau} \right) = \ii K (\tau) = \ii K (k) \; .
\ee
Moreover,
\bea
  \sn (u; k') = - \ii { {\rm sn} (\ii u ; k) \over {\rm cn} ( \ii u ; k ) } \; , & \qquad &
  \cn (u; k') =  { 1 \over {\rm cn} ( \ii u ; k ) } \; , 
  \nonumber \\
  \dn (u; k') =  { {\rm dn} (\ii u ; k) \over {\rm cn} ( \ii u ; k ) } \; , & \qquad & \ldots
\eea
   
For the ${\bf T}$ transformation we have
\begin{subequations}
\bea
 \theta_1 \left( z |  \tau + 1 \right) &= & \eu^{\ii \pi/4} \theta_1 ( z | \tau) \; , \\
 \theta_2 \left( z |  \tau + 1 \right) &=& \eu^{\ii \pi/4} \theta_2 ( z | \tau) \; ,\\
 \theta_3 \left( z |  \tau + 1 \right) &=&  \theta_4  ( z | \tau) \; , \\
\theta_4 \left (z |  \tau + 1 \right) &=&  \theta_3 ( z | \tau) \; ,
\eea
\label{TthetaTrans}
\end{subequations}
and thus
\be 
  k ( \tau +1 ) = \ii {k (\tau) \over k^\prime (\tau) } \; , \qquad \qquad \quad
  k^\prime (\tau + 1 ) =  {1 \over k^\prime (\tau) } \; ,
\ee
and
\be
  K (\tau +1 ) = k^\prime (\tau) K (\tau) \; , \qquad \quad
  \ii K^\prime (\tau + 1) =  k^\prime \left[ K (\tau) + \ii K^\prime (\tau) \right] \; .
\ee
Finally
\bea
  \sn \left( u; \ii {k \over k'} \right) =  k- { {\rm sn} (u/k' ; k) \over {\rm dn} ( u/k' ; k ) } \; , & \quad &
  \cn  \left( u; \ii {k \over k'} \right)  =  { {\rm cn} (u/k' ; k) \over {\rm dn} ( u/k' ; k ) }  \; ,
  \nonumber \\
  \dn  \left( u; \ii {k \over k'} \right)  =  { 1 \over {\rm dn} ( u/k' ; k ) } \; , & \quad & \ldots
\eea

While it is true that by composing these two transformations we can generate the whole group, for convenience we 
collect here the formulas for another transformation we use in the body of the paper: $\tau \to { \tau \over 1 - \tau}$, 
corresponding to ${\bf STS}$
\begin{subequations}
\bea
 \theta_1 \left( z |  {\tau \over 1 - \tau} \right) &=&  \ii^{\frac 1 2} F \: \theta_1 \Big ( (1 - \tau) z | \tau \Big) \; , \\
 \theta_2 \left( z |  {\tau \over 1 - \tau} \right) &=&  F \: \theta_3 \Big ( (1 - \tau) z | \tau \Big) \; , \\
 \theta_3 \left( z  | {\tau \over 1 - \tau} \right) &=&  F \: \theta_2 \Big ( (1 - \tau) z | \tau \Big) \; , \\
 \theta_4 \left (z |  {\tau \over 1 - \tau} \right) &=&  \ii^{\frac 1 2} F \: \theta_4 \Big ( (1 - \tau) z | \tau \Big) \; ,
\eea
\label{SthetaTrans}
\end{subequations}
where $F = (1 - \tau)^{1 \over 2} \eu^{\ii {(\tau -1) z^2 \over \pi}}$.
We have
\be
   k \left( {\tau \over 1- \tau} \right) = {1 \over k ( \tau )} = {1 \over k} \; , \qquad \qquad \quad
   k^\prime  \left( {\tau \over 1 - \tau} \right) = \ii {k^\prime ( \tau ) \over k (\tau)} = \ii {k' \over k} \; ,
\ee
and
\be
  K \left( {\tau \over 1 - \tau} \right) = k \left[ K ( \tau) - \ii K^\prime (\tau) \right]  \; , \qquad 
  \ii K^\prime \left( {\tau \over 1 - \tau} \right) = \ii k K^\prime (\tau) \; .
\ee
Moreover,
\bea
  \sn \left( u; {1 \over k} \right) = k \: {\rm sn} \left ( {u \over k} ; k \right) \; , & \qquad &
  \cn \left( u; {1 \over k} \right) =  {\rm dn}  \left ( {u \over k} ; k \right) \; , 
  \nonumber \\
  \dn \left( u; {1 \over k} \right) =  {\rm cn}  \left ( {u \over k} ; k \right) \ \; , & \qquad & \ldots
  \label{jacobiSTS}
\eea

\section*{References}{}

\end{document}